\documentclass[aps,showpacs,preprint]{revtex4}
\usepackage{amsmath}
\usepackage{epsfig}

\begin{document}
\title{The Transverse Electron Scattering Response Function of $^3$He }

\author{Sara Della Monaca$^{1}$, Victor D. Efros$^{2}$, Avas Khugaev$^{3}$,
  Winfried Leidemann$^{1,4}$, 
  Giuseppina Orlandini$^{1,4}$, 
  Edward L. Tomusiak$^{5}$,
  and Luping P. Yuan$^1$}
\affiliation{$^{1}$Dipartimento di Fisica, Universit\`a di Trento,
  I-38100 Trento, Italy\\
$^{2}$ Russian Research Centre "Kurchatov Institute",
  123182 Moscow,  Russia\\
$^{3}$Institute of Nuclear Physics, Uzbekistan Academy of Sciences,
  Tashkent, Uzbekistan\\
$^{4}$Istituto Nazionale di Fisica Nucleare, Gruppo Collegato
  di Trento, Italy\\
$^{5}$Department of Physics and Astronomy
  University of Victoria, Victoria, BC V8P 1A1, Canada\\
}

\date{\today}

\begin{abstract}
The transverse response function $R_T(q,\omega)$ for $^3$He
is calculated using the configuration space BonnA nucleon-nucleon potential,
the Tucson-Melbourne three-body force, and the Coulomb potential.
Final states are completely taken into account via the Lorentz integral
transform technique.
Non-relativistic one-body currents plus two-body $\pi$- and $\rho$-meson 
exchange currents as well as the Siegert operator are included.  
The response $R_T$ is calculated for $q$=174, 250, 400, and 500 MeV/c
and in the threshold region at $q$=174, 324, and 487 MeV/c. 
Strong MEC effects are found in low- and high-energy tails, but due to MEC
there are also moderate enhancements of the quasi-elastic peak (6\%-10\%). 
The calculation is performed both directly and via transformation
of electric multipoles to a form that involves the charge operator. 
The contribution of the latter operator
is suppressed in and below the quasielastic peak while at higher energies
the charge operator represents almost the whole MEC contribution at the lowest $q$ value. 
The effect of the Coulomb 
force in the final state interaction is investigated for the threshold region 
at $q$=174 MeV/c. Its neglect enhances $R_T$ by more than 10\% in the range up to
2 MeV above threshold.
In comparison to experimental data one finds relatively good agreement at $q$=250 and 400
MeV/c, while at $q$=500 MeV/c, presumably due to relativistic effects, the theoretical 
quasi-elastic peak position is shifted to somewhat higher energies. The strong MEC 
contributions in the threshold region are nicely confirmed by data at $q$=324 and 487 MeV/c.

\end{abstract}

\bigskip

\maketitle
\section{Introduction}

Electromagnetic interactions in the trinucleon systems play
an important role in testing
NN and 3N forces as well as nucleonic current operators.  Among
the many reaction observables available are the response
functions which determine the inclusive electron scattering
cross-section
\begin{eqnarray}
\frac {d^2\sigma}{d\Omega\,d\omega}\ =\ \sigma_{Mott}\ \bigg[\ \frac {Q^4}{q^4}\,
R_L(q,\omega)\ +
\ \left(\ \frac{Q^2}{2q^2}+\tan^2 \left(\frac{\theta}{2}\right)\ \right)\,R_T(q,\omega)\bigg]
\end{eqnarray}
where $\omega$ is the electron energy loss, $q$ is the magnitude of the
electron momentum transfer, $\theta$ is the electron scattering angle and
$Q\equiv\{{\bf q},\omega\}, Q^2=q^2-\omega^2$. $R_L(q,\omega)$ and $R_T(q,\omega)$ are called
the longitudinal and transverse response functions respectively.
In order to calculate either of these response functions one needs to
be able to take into account all final states (usually in the
continuum) which are connected to the ground state via the current or charge operators.
This can be accomplished in several ways. For example recent calculations
of these response functions by Golak~{\it et al} \cite{Golak05} and 
Deltuva~{\it et al} \cite{Deltuva1} base their calculations
on Faddeev techniques while we employ the Lorentz integral transform (LIT)
\cite{ELO94,ELOT04} method. A recent review article on the LIT approach is given
in \cite{ELOB07}.

The longitudinal response is driven by the nuclear charge
density operator and has recently been calculated covering large parts of the
non-relativistic regime in Refs. \cite{Golak05,ELOT04,Deltuva1}.
It is notable that although some of these groups use 
considerably different calculational techniques 
they obtain similar results for $R_L(q,\omega)$.
All these non-relativistic calculations of $R_L(q,\omega)$
show quite good agreement with experiment for modest momentum transfers
i.e. $q<400$ MeV/c.  However for larger $q$ the position of the quasi-elastic 
peak is sensitive to relativistic corrections in the kinetic energy. 
In \cite{ELOT05} the non-relativistic calculation was extended up to $q$=700 MeV/c 
by choosing a proper reference frame, where relativistic effects on the kinetic
energy are minimized. A subsequent transformation of the theoretical
results to the laboratory frame led in fact to a much better agreement with data.
Concerning the realistic NN interaction model there appears to be a relative 
insensitivity to which model is used. 
As far as 3N forces are concerned there is no unique picture. For $^3$He
their inclusion improves the agreement with data, whereas for $^3$H one
observes the opposite effect \cite{ELOT04}.

In $R_T(q,\omega)$ it is the nuclear transverse current density which drives
the response.
This current density can be expressed as the sum of
various components: the normal one-body currents with their relativistic
corrections, two-body currents
arising from meson exchange NN forces (MEC) and isobar excitations,
three-body currents arising from NNN forces. 
In \cite{Golak05} the AV18 NN potential \cite{AV18} with the UrbanaIX NNN potential 
\cite{UrbIX} were used and one-body currents as well as $\pi$- and $\rho$-MEC were 
taken into account to calculate $R_T(q,\omega)$ at q=200, 300, 400, and 500 MeV/c 
and the low-energy $R_T$ at various $q$. In \cite{Deltuva1} the CD-Bonn potential and 
its coupled channel extension CD-Bonn+$\Delta$ \cite{CDB} was taken.  The one-body 
current, $\pi$- and $\rho$-MEC and $\Delta$-currents were considered.
In addition to computing $R_T$ at q=300 and 500 MeV/c  near threshold responses
at various $q$ were also shown. 

Here we present the first fully realistic computation of $R_T$ with the
LIT method (in \cite{Bacca} the LIT method was applied to the $R_T$ of $^4$He but 
with approximations for the MEC and using a semirealistic NN potential only). 
We use the configuration space BonnA potential \cite{BonnRA} 
(hereinafter referred to as BonnRA)
together with the TM' \cite{TM} NNN potential to calculate the response
at q=174, 250, 400, 500 MeV/c and at q=324 and 487 MeV/c in the near
threshold region. Our reason for choosing the BonnRA pontential 
is that the MECs are uniquely defined in the case of a boson exchange based potential.
We would like to emphasize that the LIT method allows us to include consistently the Coulomb
interaction in initial and final states, which is not done in \cite{Golak05,Deltuva1}.
For the electromagnetic current operator we include the non-relativistic
one-body operators plus, as in \cite{Golak05,Deltuva1}, the $\pi$ and $\rho$ two-body MEC
currents. As is well known the MEC are intimately connected to details
of the Hamiltonian through the requirement of charge conservation.
With boson-exchange potentials like the BonnRA and partially for CD-Bonn 
(contains two effective $\sigma$-mesons with partial-wave dependent
parameters) the form of the MEC are determined by explicit knowledge of the 
boson-nucleon coupling.
For phenomenological potentials such as the AV18 which was used in \cite{Golak05} 
one can construct a consistent $\pi$- and $\rho$-MEC \cite{Riska,Buchmann,ArenSchwamb} 
by interpreting the isovector part of the given potential model as due to an effective
$\pi$ and $\rho$ exchange.

Our calculation is performed in two ways depending on how we treat the electric
multipole operators.  One method, which we refer to as the direct method, simply
uses the current operators {\it per se} in the electric multipoles. In the
second method the electric multipole operators are transformed via use of the
continuity equation into a form which includes the charge operator. We refer
to this latter form of the electric multipole operator as the Siegert form.
If both the continuity equation were fulfilled exactly and dynamic equations
were solved exactly then these two ways would lead to the same results.
Since, as in our case, a realistic nuclear force includes components additional 
to one-boson exchange
potentials, such as momentum-dependent NN forces and 3N forces, the continuity 
equation is only approximately fulfilled when one employs only the dominant, well established 
MECs. Therefore performing calculations in the two ways  allows us on  
 one hand to 
find out to what extent the $\pi$ and $\rho$ exchange currents we use
are compatible with the realistic nuclear force employed.
On the other hand,  via use of the charge operator 
it permits us to take into account a part of the 
additonal MEC  
thus checking their possible relevance. In \cite{Golak05,Deltuva1} such
an investigation has not been carried out (in \cite{Golak05} the Siegert operator is only used
for reactions with real photons).

\section{Nuclear forces and the current operator} 

The transverse response $R_T$ which
depends on the transverse nuclear current density operator
${\bf J}_T$ is given by
\begin{equation}
R_T(q,\omega)=\overline{\sum}_{M_0}\sum\!\!\!\!\!\!\!\!\int df\langle\Psi_0|
{\bf J}_T^\dag({\bf q},\omega)|\Psi_f\rangle\cdot\langle \Psi_f|{\bf J}_T({\bf q},\omega)|
\Psi_0\rangle\,\,\delta(E_f-E_0+q^2/(2M_T)-\omega).\label{rl}
\end{equation}    
Here $M_T$ is the mass of the target nucleus, $\Psi_0$ and $\Psi_f$ denote the 
ground and final states, respectively, while $E_0$ and $E_f$ are their 
eigenenergies,  
\begin{equation}
(h-E_0)\Psi_0=0,\qquad (h-E_f)\Psi_f=0,
\end{equation}
where $h$ is the intrinsic nuclear non-relativistic Hamiltonian.
States of our system are represented by products of
normalized center of mass plane waves $\varphi({\bf P}_{0,f})$ and internal substates
$\Psi_{0,f}$ entering (\ref{rl}). Correspondingly, the current
operator $\bf J$ in (\ref{rl}) 
is related to
 the primary current operator ${\bar{\bf J}}$ as follows,
 \begin{equation}
 {\bf J}\,\delta({\bf P}_f-{\bf P}_0-{\bf q})=
 \langle\varphi({\bf P}_{f})|{\bf {\bar J}}|\varphi({\bf P}_{0})\rangle,\label{def}
 \end{equation}
 where the matrix element is defined in the center of mass subspace.
 The cross section we need corresponds to the laboratory reference
 frame and we set in (\ref{def}) ${\bf P}_0=0$.
 The quantity ${\bf J}_T$ is that component of ${\bf J}$ which is
 orthogonal to ${\bf q}$. The second summation (integration) in (\ref{rl})
 goes over all final states belonging to the same
 energy $E_f$, and $M_0$ is the projection of the ground state angular momentum $J_0$.

The Hamiltonian $h$ includes the kinetic energy terms, the 2N and 3N force terms, 
and the proton Coulomb interaction term. As in \cite{ELOT04} the ground state 
$\Psi_0$ is calculated via an expansion in basis functions which are correlated 
sums of products of hyperradial functions, hyperspherical harmonics and 
spin-isospin functions. In the present work the 2N + 3N interactions are
taken as the Coulomb+ BonnRA+TM$^\prime$ ($\Lambda$=2.835 fm$^{-1}$) as in
\cite{ELOT04}.
The TM$^\prime$ cut-off parameter $\Lambda$ 
properly fixes the $^3$H binding energy to 8.47 MeV. 

We perform a non-relativistic calculation. The current ${\bf J}$ includes
one-body and two-body operators. The one-body current operator as obtained
from (\ref{def}) is
\begin{equation}
{\bf {j}}^{(1)}=\ \sum_{k=1}^A\ [\ {\bf j}(k)_{\rm spin}+{\bf j}(k)_p+{\bf j}(k)_q \  ] \nonumber
\end{equation}
where $A$ is the number of nucleons in the target nucleus and
\begin{eqnarray}
{\bf j}(k)_{\rm spin}=\
e^{i{\bf q}\cdot{\bf r}_k^\prime}\ \frac{i({\sigma_k}\times{\bf q})}{2M}\
G_M(k),\nonumber\\
{\bf j}(k)_{p}=\ e^{i{\bf q}\cdot{\bf r}_k^\prime}\ \frac{{\bf p}_k^\prime}{M}\  G_E(k)
,\nonumber\\
{\bf j}(k)_{q}=\ e^{i{\bf q}\cdot{\bf r}_k^\prime}\ \frac{{\bf q}}{2M}\  G_E(k). \nonumber
\end{eqnarray}
%
Here ${\bf r}_k^\prime = {\bf r}_k-{\bf R}_{cm}$, 
${\bf p}_k^\prime = {\bf p}_k-{\bf P}_{cm}/A$, and ${\sigma}_k$ are the relative 
coordinate, momentum, and spin operator of the k-th particle and $M$ denotes the 
nucleon mass, while ${\bf R}_{cm}$ and ${\bf P}_{cm}$ are the center of mass 
coordinate and momentum variables of the $A$-body system. The component ${\bf j}_{q}$
does not contribute to ${\bf J}_T$. However, separate multipoles as defined
below depend on this component. 

In the above expressions we use the notation
\begin{equation}
G_{E,M}(k)=G_{E,M}^p(Q^2)\frac{1+\tau_{zk}}{2}+G_{E,M}^n(Q^2)\frac{1-\tau_{zk}}{2}
\end{equation}
where $G_{E,M}^{p,n}$ are the Sachs form factors and $\tau_{zk}$ 
denotes the third component of the isospin operator of the k-th 
nucleon. With our procedure the computational 
labour is reduced when the number of $\omega$-dependent form factors is reduced 
\cite{ELOB07}. To this end we use the approximation 
\begin{equation}
\label{GEn}
G_E^n(Q^2)\approx G_E^p(Q^2)\gamma(Q_{av}^2)
\end{equation}
where $\gamma(Q_{av}^2)=G_E^n(Q_{av}^2)/G_E^p(Q_{av}^2)$,
$Q_{av}^2=q^2-\omega_{av}^2$ and $\omega_{av}=q^2/(2M)$.
Similarly for the one-body spin current we use
\begin{eqnarray}
\label{GMp}
G_M^p(Q^2)\approx{\bar\mu}_p(Q_{av}^2)G_E^p(Q^2)\qquad\qquad
{\bar\mu}_p(Q_{av}^2)=\frac{G_M^p(Q_{av}^2)}{G_E^p(Q_{av}^2)}\\
\label{GMn}
G_M^n(Q^2)\approx{\bar\mu}_n(Q_{av}^2)G_E^p(Q^2)\qquad\qquad
{\bar\mu}_n(Q_{av}^2)=\frac{G_M^p(Q_{av}^2)}{G_E^p(Q_{av}^2)} .
\end{eqnarray}
For the usual dipole magnetic form factors, as used in this work, the above relations 
are fulfilled exactly and we have checked that the approximation provides a very good 
accuracy for $G^n_E$. In a future extension of our work to a high-$q$ region, $q>500$ 
MeV/c, we will use more sophisticated nucleon form factor fits. In these cases the above 
relations are only approximately fulfilled although we have checked that they still lead 
to excellent accuracy. The neutron electric form factor we use here is taken from
\cite{Platchkov} as used in \cite{Bosted}.  
With (\ref{GEn}-\ref{GMn}) the one-body current is replaced by
${\bf j}^{(1)} \to G_E^p(Q^2){\bf J}^{(1)}$ where ${\bf J}^{(1)}$ is now given by 
\begin{eqnarray}
{\bf J}^{(1)}({\bf q})=\sum_{k=1}^A\,\frac{e^{i{\bf q}\cdot{\bf
r}_k^\prime}}{M}\left\{\left({\bf p}_k^\prime+\frac{\bf q}{2}\right)\left[
\frac{1+\tau_{zk}}{2}+\gamma(Q_{av}^2)\frac{1-\tau_{zk}}{2}\right]\right.\nonumber\\
+\left.\frac{i(\sigma_k\times\bf q)}{2}\left[
{\bar\mu}_p(Q_{av}^2)\frac{1+\tau_{zk}}{2}+{\bar\mu}_n(Q_{av}^2)\frac{1-\tau_{zk}}{2}\right]\right\}.
\end{eqnarray}

The dominant contributions to the two-body current ${\bf J}^{(2)}$ arise from
the $\pi$- and $\rho$-meson exchange currents. These currents
are usually expressed in terms of  "Seagull" and "true exchange"
pieces.  Thus we write here
\begin{equation}
\label{J2}
{\bf J}^{(2)}\ =\ {\bf j}_{SG}^\pi\ +\ {\bf j}_{ex}^\pi\ +\ 
{\bf j}_{SG}^\rho\ +\ {\bf j}_{ex}^\rho .
\end{equation}
We list in Appendix A the coordinate space representations of these currents with
the corresponding values of coupling constants etc.
Momentum space forms of these meson exchange currents are related to these coordinate
space forms, apart from the multiplicative isovector electric form factor
$G_E^v(Q^2)=(G_E^p(Q^2)-G_E^n(Q^2))/2$, via 
\begin{equation}
{\bf j}_a^b({\bf q})e^{i{\bf q}\cdot{\bf R}_{cm}}
 = \int d^3x\,e^{i{\bf q}\cdot{\bf x}}\,{\bf j}_a^b({\bf x})
\end{equation}
where the super/sub-scripts above are those corresponding to the right hand 
side of (\ref{J2}).

Finally we use the current operator ${\bf J}$ in the form
\begin{equation} 
{\bf J}\ =\ G_E^p(Q^2){\bf J}^{(1)}\,+\,2G_E^v(Q^2){\bf J}^{(2)} .
\end{equation}

\section{Multipole Expansion of the Transverse Response}

The dynamic calculations are performed in separate
subspaces belonging to fixed angular momentum $J$ and
its projection $M$ (see also \cite{ELOB07}). 
One can account for $M$-dependencies analytically 
via  performing a multipole expansion of $R_T$. 
To this end we use a decomposition into multipoles of  the
transverse current. This decomposition shall also allow us 
employing an alternative expression for the transition operator, see below.
The transverse current is represented as   
\begin{equation}
{{\bf J}}_T=4\pi\sum_{\lambda={\rm el,mag}}
\sum_{jm}i^{j-\epsilon}\,{\cal T}_{jm}^{\lambda}(q){\bf Y}_{jm}^{(\lambda)*}({\hat {\bf q}}).
\end{equation}
Here ${\hat {\bf q}}=q^{-1}{\bf q}$ and ${\bf Y}_{jm}^{(\lambda) }$ are electric and magnetic
vector spherical harmonics \cite{varsh} and $\epsilon$=0 when $\lambda={\rm el}$
or $\epsilon$=1 when $\lambda={\rm mag}$. This then allows the
transverse response to be written as
\begin{equation}
\label{rcomp}
R_T(q,\omega)=
\frac{4\pi}{2J_0+1}\sum_{\lambda={\rm el,mag}}\sum_{Jj}(2J+1)(R_T)_J^{j\lambda}
\end{equation}
where 
\begin{equation}
\label{mexp}
(R_T)_J^{j\lambda}=\sum\!\!\!\!\!\!\!\int\,df \langle q_{JM}^{j\lambda}|\Psi_f(J,M)\rangle
\langle\Psi_f(J,M)|q_{JM}^{j\lambda}\rangle\delta(E_f-E_0-\omega),
\end{equation}
$J$ and $M$ are the final state angular momentum and its projection,
and $|q_{JM}^{j\lambda}\rangle$ is given by
\begin{equation}
\label{q}
|q_{JM}^{j\lambda}\rangle = [{\cal T}_j^\lambda \otimes |\Psi_0(J_0)\rangle]_{JM} .
\end{equation}
In Eq.~(\ref{mexp}) $M$ is arbitrary. 

In terms of the more standard multipoles and vector spherical harmonics
${\bf Y}_{jm}^l $ we can write
\begin{equation}
\label{Tel1}
{\cal T}_{jm}^{\rm el}=\left(\frac{j+1}{2j+1}\right)^{1/2}
{T}_{jm}^{j-1} +\left(\frac{j}{2j+1}\right)^{1/2}{T}_{jm}^{j+1},
\end{equation}
\begin{equation}
\label{Tel2}
{\cal T}_{jm}^{\rm mag} \equiv {T}_{jm}^{j}
\end{equation}
where
\begin{eqnarray}
T_{jm}^{l}=\frac{1}{4\pi i^{j-\epsilon}}\int d\Omega_{\bf q}
\left({\bf Y}_{jm}^l({\bf \hat{q}})\cdot{{\bf J}}({\bf q},\omega)\right).
\end{eqnarray}
Since charge has to be conserved  
it is well known that the above
expression for ${\cal T}_{jm}^{el}$ can be rewritten as
\begin{equation}
\label{Sieg}
{\cal T}_{jm}^{\rm el}=\left(\frac{j+1}{j}\right)^{1/2}
\frac{\omega}{q}\rho_{jm}+
\left(\frac{2j+1}{j}\right)^{1/2}
T_{jm}^{j+1}
\end{equation}
where
$\rho_{jm}$ is a charge multipole of the charge density operator $\rho$
defined by
\begin{equation}
\rho_{jm}(q)=\frac{1}{4\pi i^j}\int d\Omega_{\bf q}\ Y_{jm}(\hat {\bf q}) \rho({\bf q}) .
\end{equation}
We shall refer to the form of ${\cal T}_{jm}^{\rm el}$ in (\ref{Tel1}) as the direct form 
and to that in (\ref{Sieg}) as the Siegert form. 
The first term of (\ref{Sieg}) will be called Siegert operator, while
the second term is the residual term.
Appendix B gives the multipole operators $T_{jm}^l$
for the one-body currents while
Appendix C lists them 
for the $\pi$ and $\rho$ exchange currents. 

\section{Calculation of the Response}
The techniques we use in calculating the response have been largely
set out in \cite{ELOT04}.  Here we add some extra detail
which arises in the case of the transverse response. The Lorentz transform of 
the partial response $(R_T)_J^{j\lambda}$ of Eq.~(\ref{mexp}) is given by
\begin{equation}
\label{phi}
\Phi_{J}^{j\lambda,\alpha}(q,\sigma_R,\sigma_I)=\sum_n
\frac{(R_T)_{J}^{j\lambda,\alpha}(q,\omega_n)}
 {(\omega_n-\sigma_R)^2+\sigma_I^2}+
\int d\omega \frac{(R_T)_{J}^{j\lambda,\alpha}(q,\omega)}{(\omega-\sigma_R)^2+\sigma_I^2}.
\end{equation}
The sum in (\ref{phi}) corresponds to transitions to discrete
levels with excitation energy $\omega_n$. In our A=3 case there exists only one 
discrete contribution corresponding to M1 elastic scattering. 
In (\ref{phi}) the response is supplied with an additional superscript $\alpha$. It 
specifies separate contributions to the response $(R_T)_{J}^{j\lambda}$ of 
Eq.~(\ref{mexp}), e.g. a given $\alpha$ determines the isospin of the final state.  
In addition it specifies contributions that correspond to components
of the multipole operators with different nucleon form factor dependencies.

It was pointed out above that 
one-body and two-body currents have different $\omega$-dependence
through their different form factors. Therefore  we need to calculate the
responses with the individual parts of the current i.e. 
${\bf J}^{(1)}{\bf J}^{(1)}, {\bf J}^{(1)}{\bf J}^{(2)}$, and
${\bf J}^{(2)}{\bf J}^{(2)}$. The corresponding partial response functions will
carry the superscripts $\alpha=\{11, 12, 22\}$ so that the response $(R_T)_J^{j\lambda}$
would be expressed as
\begin{eqnarray}
\label{RTdirect}
\nonumber
(R_T(q,\omega))_J^{j\lambda}\ =&\ (G_E^p(Q^2))^2(R_T(q,\omega))_J^{j\lambda,11}
+4G_E^p(Q^2) G_E^v(Q^2) (R_T(q,\omega))_J^{j\lambda,12} \\ &+
4(G_E^v(Q^2))^2 (R_T(q,\omega))_J^{j\lambda,22} \,.
\end{eqnarray}

Additional $\omega$-dependence of the electric multipole operators arises when 
they are used in the Siegert form, i.e. in the form of Eq.~(\ref{Sieg}).
Due to the additional  $\omega$-dependence of the first term in (\ref{Sieg})
we calculate separately the response originating from this term $(\sim\omega^2)$, the response
originating from the second term in (\ref{Sieg}) and the cross-term response $(\sim\omega)$.
For the same reason as in 
Eq.~(\ref{RTdirect}) each of these responses is in turn broken into 
the one-body piece, the two-body piece
and the cross piece which are calculated separately. The superscript $\alpha$ in (\ref{phi})
enumerates all these various cases, so that the response $(R_T)_J^{j,{\rm el}}$ 
is a sum of the responses
$(R_T)_J^{j\lambda,\alpha}$ multiplied by products of nucleon form factors 
times $\omega^n$, $n=0,1,2$.

As described in \cite{ELO94} the transforms are determined dynamically. In the present case
the transforms $\Phi_J^{j\lambda,\alpha}$ are obtained from
\begin{equation}
\label{dy}
\Phi_{J}^{j\lambda,\alpha}(q,\sigma_R,\sigma_I)=\langle{\tilde \psi_{JM}^{j\lambda,\alpha}
}|{\tilde \psi_{JM}^{j\lambda,\alpha}}\rangle,
\qquad
|{\tilde \psi_{JM}^{j\lambda,\alpha}}\rangle=
[h-\sigma_R+i\sigma_I]^{-1}|q^{j\lambda,\alpha}_{JM}\rangle.
\end{equation}
The calculation (\ref{dy}) is $M$-independent and is performed in separate subspaces belonging
to given isospin and parity.  Parities are determined by the multipole order $j$ and the
choice of $\lambda$=el/mag. For a given $\lambda$, parity, and $J$ only one value of 
$j$ is possible in our case.

To pass to responses one needs to invert the transforms. This may be done either separately for
each transform $\Phi_{J}^{j\lambda,\alpha}$ using Eq.~(\ref{phi}) or for their sums 
at the same $\alpha$.
One may define the responses $R_T^\alpha=\sum_{\lambda={\rm el,mag}}R_T^{\lambda, \alpha}$,
where (c.f. (\ref{rcomp}))
\begin{equation}
\label{rcomp1}
R_T^{\lambda,\alpha}(q,\omega)=
\frac{4\pi}{2J_0+1}\sum_{Jj}(2J+1)(R_T)_J^{j\lambda,\alpha}(q,\omega).
\end{equation} 
One also defines the corresponding transforms
\begin{equation}
\label{rcomp2}
\Phi^{\lambda,\alpha}(q,\sigma_R,\sigma_I)=
\frac{4\pi}{2J_0+1}\sum_{Jj}(2J+1)\Phi_J^{j\lambda,\alpha}(q,\sigma_R,\sigma_I).
\end{equation}
They are related to the responses (\ref{rcomp1}) in the same way as in (\ref{phi}). 
All the various $\Phi^{\lambda,\alpha}$  are inverted separately to get $R_T^{\lambda,\alpha}$.
Our inversion method and more information concerning the inversion can be found in
\cite{ELOB07,ELO99,Andreasi}. We take $\sigma_I=20$ MeV and distinguish
between the two isospin cases $T$=1/2 and 3/2, since
the corresponding responses have different thresholds and thus the
inversion can be carried out more precisely. After having inverted both cases
we sum up the two results. For the magnetic part of the response we invert
the M1 transition to the final state with $J^\pi={\frac{1}{2}}^+$ separately,
since, as mentioned above, it contains an elastic contribution. This elastic
contribution can easily be determined by choosing a very small value for $\sigma_I$ 
and thereafter its effect on the transform can be subtracted leading to a
LIT of a purely inelastic response. 

As mentioned earlier charge conservation leads to the
equality of the Siegert and direct forms of the
transverse electric multipole operator. This provides
an important check on our procedures especially with respect to
the implementation of the MECs.  The BonnRA potential
contains more than just $\pi$- and $\rho$-meson exchange
but we expect that taking account of MECs from only
these two exchanged particles should 
lead to the dominant MEC contribution 
in our kinematical range, while additional MEC effects
are partially taken care of by the Siegert operator. 
A good test for the implementation of the MEC is provided by using 
a simple $\pi + \rho$ OBEP with their corresponding MECs. In this case
charge conservation should be exact and the transverse
response should be independent of whether one uses
the Siegert or the direct form of $T_{jm}^{el}$. 
We have made such tests at $q$=10, 300, and 500 MeV/c and found very
good agreement between the results of the two calculations \cite{dmonaca}.

\section{Results and Discussion} 

We have selected the momentum transfers $q$ = 174, 250, 400, and 500 MeV/c for a 
calculation of $R_T(q,\omega)$ in a large $\omega$-range. In addition we 
consider the low-$\omega$ part of $R_T$ at $q$ = 174, 324, and 487 MeV/c
for which cases we take a maximal value of $J=7/2$. For 
the other $q$-values a different choice for $J^{max}$ is made: 
11/2 ($q$=250 MeV/c), 15/2 ($q$=400 MeV/c), and 19/2 ($q$=500 MeV/c). We have checked 
that with these settings very good convergences of the multipole expansions 
of $R_T$ are obtained in the requested energy ranges.

In the discussion we compare results calculated with the various current operators of section
II (both with direct and Siegert forms) representing the following contributions:
(a) one-body, (b) one-body and implicit MEC via Siegert operator,
(c) one-body, $\pi$- and $\rho$-MEC, and (d) one-body,
$\pi$- and $\rho$-MEC plus additional MEC via Siegert operator.
If exact charge conservation was satisfied then the results of the direct calculation (c)
would agree with those of the
Siegert form (d).

In Figs.~1 and 2 we show the various current contributions to $R_T$. It is readily 
seen that there are rather strong MEC effects: 15-30 MeV above threshold MEC 
enhance $R_T$ by more than 30\% for the two higher $q$-values (very close to threshold 
even by up to 200\%, see Fig.~4); they increase the quasi-elastic peak height by 
10\% ($q$=174, 250 MeV/c), 7\% ($q$=400 MeV/c), and 6\% ($q$=500 MeV/c); for lower $q$ 
they also lead to large effects in the high-energy tail (e.g. at pion threshold: 
increases of 180\% ($q$=174 MeV/c), 95\% ($q$=250 MeV/c), 22\% ($q$=400 MeV/c), 
and 5\% ($q$=500 MeV/c)). 
In general, relative contributions of MEC
are determined mainly by distances $|\omega-\omega_{peak}|$. This is
natural since the peaks correspond to maximum contributions of one-body
operators.

It is also seen that Siegert contributions remain quite small in and below
the quasi-elastic peak. On the other hand they become more important with
increasing energy (e.g., at pion threshold and $q$=174 (250) MeV/c, enhancements
are of 130\% (55\%) of the one-body contribution). In addition to the fact
that in general MEC contributions  are rather small in the peak as compared to
one-body contributions, Siegert contributions are strongly suppressed in and
below the peak by the factor $\omega/q$ in (\ref{Sieg}). The approximate
transition operator we discuss takes account of MEC only via the Siegert
operator, i.e. the charge operator from (\ref{Sieg}). As it is seen from
Fig.~2, in the tail region  this approximation provides the response rather
close to the true one at the lowest $q$ value $q=174$ MeV/c. This agrees with
the well-known fact that in moderate energy photodisintegration processes ($\omega=q$) 
MEC contributions are largely included by the Siegert operator.

It is interesting to note that there exist additional Siegert MEC contributions
beyond the $\pi$- and $\rho$-MEC entering the direct calculation.
This is due to the fact that $\pi$ and $\rho$ exchanges constitute only the 
dominating part of a consistent exchange current with the BonnRA potential. 
Other two-body currents are induced by momentum and spin-orbit dependent 
potential terms. In addition also three-body currents, originating
from the TM-3NF, could lead to Siegert contributions.
Effects of the Siegert operator beyond $\pi$- and $\rho$-MEC
were also found in the proton-deuteron radiative capture with
the BonnCD+$\Delta$ potential \cite{Deltuva2004}.
It is seen from Fig.~1 and Fig.~2 that such effects are small
for energies far away from the photon point. Indeed, there the complete 
calculation via direct inclusion of MEC operators and the complete alternative calculation 
that involves the Siegert operator have led to results close to each other.
However, closer to the photon point the additional Siegert
contributions can lead to corrections of the order of 10\%. 

In Fig.~3 we show our $R_T$ results in comparison to experimental data. For 
$q$= 250 and 400 MeV/c one finds good agreement. However, data are not precise
enough to allow a definite conclusion about the MEC contribution. As opposed
to the lower $q$ cases we find at $q$=500 MeV/c a difference between the
theoretical and experimental peak positions.
The shift amounts to about 5-10 MeV. Relativistic effects, 
in particular those arising from corrections to the kinetic energy,
might be responsible for this difference. In fact in \cite{ELOT05} it was
shown for the longitudinal response function $R_L(q,\omega)$ that such
effects lead at $q$=500 MeV/c to a shift of the peak position  by 6 MeV.

In Fig.~4 we depict various $R_T$ theoretical and experimental low-energy results
at $q$=0.882, 1.64, and 2.47 fm$^{-1}$ corresponding to about 174, 324, and 487
MeV/c, respectively.
We do not show the contribution of the Siegert operator, since, as shown in Fig.~1,
its effect is very small at low energies.
One sees that the MEC contribution can be very important, e.g. at q=487 MeV/c one finds
an increase of about 200\% close to threshold.
Contrary to the cases shown in Fig.~3 one can make a definite conclusion
about the MEC contribution. It is evident that they lead to a considerably 
improved agreement between theory and experiment. For the two higher $q$-values 
theoretical and experimental results agree very well, whereas for $q$=174 MeV/c the 
theoretical result underestimates experimental data somewhat below 10 MeV. 
A better theoretical description of the $q$=174 MeV/c data is found in 
\cite{Golak05}, where the AV18 NN potential and the UrbanaIX 3N-force is used as 
nuclear interaction.  However, the Coulomb force was not included in the 
final state interaction. The effect of such a neglect is illustrated in Fig.~4 
for the case in discussion. Within 2 MeV above threshold it leads to an 
increase of more than 10\%, while at 5 MeV above threshold the effect still amounts to 4\%.
In this way the theoretical results are shifted closer 
to the experimental data, but the effect is too small to reach a good agreement 
at low energies.

We summarize our results as follows. We have calculated the transverse form 
factor $R_T(q,\omega)$ considering besides one- and two-body currents
also the so-called Siegert operator.
As nuclear interaction we have taken the BonnRA NN potential and the 
Tucson-Melbourne TM' 3N-force. Since we are particularly interested in
the MEC effects and the role of the Siegert operator we have chosen the BonnRA 
potential, for which the important $\pi$- and $\rho$-exchange currents are 
directly determined by the potential model. It is true that also for more 
phenomenological NN potentials, e.g. AV18, a consistent $\pi$- and $\rho$-MEC 
can be constructed \cite{Riska,Buchmann,ArenSchwamb}, but to this end one has 
to interpret the isovector part of the phenomenological potential as an 
effective $\pi$- and $\rho$-exchange. 

We find that MEC provide very strong contributions both at lower energies and
in the high-energy tail while giving a moderate increase to the height
of the quasi-elastic peak. Siegert contributions are unimportant
in and below the quasi-elastic peak. They become considerably more sizeable at 
higher energies, but additional MEC contributions have also to be taken into account
and thus a calculation, where in addition to the one-body current, MEC currents are
taken into account via only the Siegert operator is not sufficient.
On the other hand also a calculation with only
one-body currents plus $\pi$- and $\rho$-MEC may not be  sufficient at higher 
energies, since, as we have shown, effects due to additional two- and three-body 
currents can become important. To include at least a part of these additional 
exchange effects it is better to work also in this case with the Siegert 
operator. The appropriate place
to study the structure of MEC is the energy region below the quasielastic
peak. Indeed, the contributions of MEC are large in this region while
they are rather small in the peak, and beyond the peak they are partly
represented by the Siegert operator, i.e. the charge operator.

In comparison to experiment relatively good agreement is obtained 
at $q$=250 and 400 MeV/c, while at 
$q$=500 MeV/c, presumably because of relativistic effects, the position of the 
theoretical quasi-elastic peak is located somewhat above the experimental one.
Close to threshold one finds very strong MEC contributions. They are necessary in order
to achieve a good description of the experimental data at q=324 and 487 MeV/c.
Also at $q$=174 MeV/c they lead to an improved agreement with experiment, but 
in the range from threshold to 5 MeV above the theoretical result underestimates data
somewhat.

In future we plan to investigate the momentum range 500 MeV/c $\le q \le$ 1 GeV/c
considering relativistic corrections for the one-body current operator and performing
the calculation in a reference frame where relativistic effects
in the kinetic energy are minimized \cite{ELOT05}. We also plan to study isobar
current contributions including $\Delta(1232)$ degrees of freedom.

\section{Acknowledgment}

Acknowledgements of financial support are given to
the Russian Foundation for Basic Research, grant 07-02-01222-a (V.D.E.)
and to the National Science and Engineering Research Council of Canada (E.L.T.).

\appendix

\section{Configuration Space $\pi$ and $\rho$ MECs}

For convenience we list below the well known $\pi$ and $\rho$ configuration
space exchange currents. 
\begin{eqnarray}
{\bf j}_{SG}^{\pi}({\bf x})\ =\ \frac{f_0^2}{m_\pi^2}\,
\sum_{i<j}\,({\bf \tau}_i\times{\bf \tau}_j)_z \nonumber\\
\times\big[({\bf\sigma}_i\cdot{\bf\nabla}_i)
{\bf\sigma}_j\delta({\bf x-r}_j)\ -\ ({\bf\sigma}_j\cdot{\bf\nabla}_j)
{\bf\sigma}_i\delta({\bf x-r}_i)\big]\sum_{k=1}^3h^{\pi}_kY(\mu^{\pi}_k,|{\bf r}_i-{\bf r}_j|),\\
\nonumber\\
{\bf j}_{ex}^{\pi}({\bf x})\ =\ \frac{1}{4\pi} \frac{f_0^2}{m_\pi^2}\,
\sum_{i<j}\,({\bf \tau}_i\times{\bf\tau}_j)_z ({\bf\sigma}_i\cdot{\bf\nabla}_i)
({\bf\sigma}_j\cdot{\bf\nabla}_j) \nonumber\\
\times\sum_{k=1}^3h^{\pi}_k\big[Y(\mu^{\pi}_k,|{\bf x- r}_j|){\bf\nabla_x}Y(\mu^{\pi}_k,|{\bf x- r}_i|)-
Y(\mu^{\pi}_k,|{\bf x- r}_i|){\bf\nabla_x}Y(\mu^{\pi}_k,|{\bf x- r}_j|)\big],\\
\nonumber\\
{\bf j}_{SG}^{\rho}({\bf x})\ =\ \frac{1}{4\pi} \left(\frac{g_\rho}{2M}\right)^2
\left(1+\frac{f_\rho}{g_\rho}\right)^2\,
\sum_{i<j}\,({\bf \tau}_i\times{\bf \tau}_j)_z \nonumber\\
\times\big[({\bf\sigma}_i\times{\bf \nabla}_i)
 \times{\bf \sigma}_j  \delta({\bf x - r}_j)\,-\,
({\bf\sigma}_j\times{\bf \nabla}_j) \times{\bf \sigma}_i
\delta({\bf x - r}_i)\big]\sum_{k=1}^3h^{\rho}_k Y(\mu^{\rho}_k,|{\bf r}_i - {\bf r}_j)|).\\
\nonumber\\
{\bf j}_{ex}^{\rho}({\bf x})\ =\ \frac{1}{(4\pi)^2} \left(\frac{g_\rho}{2M}\right)^2
\left(1+\frac{f_\rho}{g_\rho}\right)^2\, \sum_{i<j}\,({\bf \tau}_i\times{\bf\tau}_j)_z\,
({\bf\sigma}_i\times{\bf\nabla}_i)\cdot({\bf\sigma}_j\times{\bf\nabla}_j)\nonumber\\
\times\sum_{k=1}^3h^{\rho}_k\big[Y(\mu^{\rho}_k,|{\bf x- r}_j|){\bf\nabla_x}Y(\mu^{\rho}_k,|{\bf x- r}_i|)-
Y(\mu^{\rho}_k,|{\bf x- r}_i|){\bf\nabla_x}Y(\mu^{\rho}_k,|{\bf x- r}_j|)\big]. 
\nonumber\\
\end{eqnarray}
Here $Y(m,r)\ =\ e^{- m r}/r$. We list the coupling constants, the  masses $\mu^{\alpha}_k$,
and the regularization constants $h^\alpha_k$, where $\alpha=\pi$ or $\rho$, taken
from \cite{BonnRA}: 
\begin{equation}
f_0^2\ =\ \frac{1}{4\pi} f_{\pi NN}^2\ =\ 0.0805, \qquad \frac{g_\rho^2}{4\pi}=1.2,
\qquad \frac{f_\rho}{g_\rho}=6.1,
\nonumber\\
\end{equation}
\begin{equation}
\label{mas}
\mu^{\alpha}_1=m_\alpha,\qquad\mu^{\alpha}_2=\Lambda_{\alpha}+10\,\,{\rm MeV},\qquad
\mu^{\alpha}_3=\Lambda_{\alpha}-10\,\,{\rm MeV},
\end{equation}
\begin{equation}
\Lambda_\pi=1.3\,\,{\rm GeV},\qquad    \Lambda_\rho=1.2\,\,{\rm GeV},
\end{equation}
\begin{equation}
\label{const}
h^{\alpha}_1=1,\qquad h^{\alpha}_2=-\frac{(\mu^{\alpha}_3)^2 - (\mu^{\alpha}_1)^2}
{(\mu^{\alpha}_3)^2-(\mu^{\alpha}_2)^2},
\qquad h^{\alpha}_3=
\frac{(\mu^{\alpha}_2)^2 - (\mu^{\alpha}_1)^2}{(\mu^{\alpha}_3)^2-(\mu^{\alpha}_2)^2}.
\end{equation}

\medskip
\section{$T_{jm}^l$ Multipoles of One-Body Currents}

In the following the non-relativistic expressions of the electric
and magnetic multipoles for the one-body currents are
written. Each of them is decomposed in a convection and a spin
current.  For the magnetic
multipoles one has
\begin{eqnarray}
{T}^{j}_{jm}=\sum_i [{T}^{j,\,{\rm spin}}_{jm}(i) + 
{T}^{j,\,{\rm conv}}_{jm}(i)]
\end{eqnarray}
with:
\begin{eqnarray}
\!\!\!\!{T}^{j,\,{\rm spin}}_{jm}(i)=\frac{1}{M}\frac{q}{2}
\left(\frac{\overline{\mu}_p+\overline{\mu}_n}{2}+
\frac{\overline{\mu}_p-\overline{\mu}_n}{2}\tau_{zi}\right)
\left\{
\sqrt{\frac{j}{2j+1}}j_{j+1}(qr'_i)[Y_{j+1}(\hat{\mathbf{r}}'_i)\otimes
\sigma_i]_{jm}\right. + \nonumber \\ \!\!\!\!\!\!\!\left.-
\sqrt{\frac{j+1}{2j+1}}j_{j-1}(qr'_i)[Y_{j-1}(\hat{\mathbf{r}}'_i)\otimes
\sigma_i]_{jm}\right\},
\end{eqnarray}

\begin{eqnarray}
{T}^{j,\,{\rm conv}}_{jm}(i)=\frac{1}{M}
\left(\frac{1+\gamma}{2}+\frac{1-\gamma}{2}\tau_{zi}\right)j_j(qr'_i)\left[Y_j(\hat{\mathbf{r}
}'_i)\otimes
\partial'_i\right]_{jm}.
\end{eqnarray}
The quantity $\partial'_\mu$ is defined by the relationship $-i\partial'_\mu=p'_\mu$.
If the last
Jacobi vector is defined as
$\vec{\xi}_{A-1}=\sqrt{(A-1)/A}\,\,
[{\bf r}_A-(A-1)^{-1}\sum_{i=1}^{A-1}{\bf r}_i]$
then
\[\partial'^{(A)}_\mu=
\left[\frac{A-1}{A}\right]^{1/2}\frac{\partial}{\partial\xi_{A-1,\mu}}.\]

Similarly we write the one-body multipoles contributing to ${\cal T}^{\rm el}_{jm}$ as 
\begin{eqnarray}
T^{l}_{jm}=\sum_i [T^{l,\,{\rm spin}}_{jm}(i) + T^{l,\,{\rm conv}}_{jm}(i)]
\end{eqnarray}
where $l=j\pm 1$.
One obtains
\begin{eqnarray}
T^{j\pm 1,\,{\rm spin}}_{jm}(i)=-\frac{1}{M}\frac{q}{2}\left(\frac{\overline{\mu}_p+\overline{\mu}_n}{2}+
\frac{\overline{\mu}_p-\overline{\mu}_n}{2}\tau_{zi}\right)\sqrt{\frac{j+(1\mp
1)/2} {2j+1}}j_j(qr'_i)\left[Y_j
(\hat{\mathbf{r}}_i')\otimes \sigma_i\right]_{jm}
\end{eqnarray}
and
\begin{eqnarray}
\label{tjconv}
T^{j\pm 1,\,{\rm conv}}_{jm}(i)=\pm\frac{1}{M}
\left(\frac{1+\gamma}{2}+\frac{1-\gamma}{2}
\tau_{zi}\right)\left\{j_{j\pm 1}(qr'_i)\left[Y_{j\pm 1}(\hat{\mathbf{r}}'_i)\otimes
\partial'_i\right]_{jm}+\right.\nonumber
\\\left.-\frac{q}{2}\sqrt{\frac{j+(1\pm
1)/2}{2j+1}}j_{j}(qr'_i)Y_{jm}(\hat{\mathbf{r}}'_i)\right\}.
\end{eqnarray}
The term proportional to $j_j(qr'_i)$ in (\ref{tjconv}) above cancels when 
one forms the electric multipole (\ref{Tel1}).

\medskip

\section{$T_{JM}^L$ Multipoles of $\pi$ and $\rho$ MECs}

Here the $T_{JM}^L$ multipoles are given for the "12" pair.  The total result
should be multiplied by 3 to account for three pairs of identical particles
in the trinucleons.

\medskip

1.{\underline {$\ \ \pi$-Seagull}}
\begin{eqnarray}
\label{c1}
T_{JM}^L\ =\ \frac{\sqrt{4\pi}}{i^{J-\epsilon}}\left(\frac{f_0}{m_\pi}\right)^2\,
\sum_{\ell \rho \sigma }
\sum_{\sigma^\prime {\cal L}}\,i^{\sigma^\prime - \ell}
(-1)^{\sigma+{\cal L}+J}[1+(-1)^{\ell+\rho}]\hat\ell^2 \hat\rho \hat L\hat\sigma
\hat\sigma^\prime\hat{\cal L}\nonumber\\
\ \left(\begin{array}{ccc}\ell &1 & \sigma \\ 0 & 0 & 0 \end{array}\right)
\left(\begin{array}{ccc}\sigma^\prime&\ell&L\\ 0&0&0\end{array}\right)
\left\{\begin{array}{ccc}L&\ell&\sigma^\prime\\ \sigma&{\cal L}&1\end{array}\right\}
\left\{\begin{array}{ccc}L&1&{\cal L}\\ \rho&J&1\end{array}\right\}\nonumber\\
j_{\sigma^\prime}(qz)\ j_\ell\left(\frac{qr}{2}\right)  
\ (H^{\pi}(r))'\
\left[[Y_{\sigma^\prime}({\hat{\bf z}})\otimes Y_\sigma({\hat{\bf r}})]^{\cal L}
\otimes \Sigma_{12}^{[\rho]}\right]_M^J\,\,(\tau_1\times\tau_2)_z\ .
\end{eqnarray}
Here ${\bf r}={\bf r}_2-{\bf r}_1$, ${\bf z}=-[({\bf r}_1+{\bf r}_2)/2-{\bf R}_{cm}]$,
${\hat{\bf r}}={\bf r}/r$, ${\hat{\bf z}}={\bf z}/z$, 
$(H^{\pi}(r))'\ =\ \ dH^\pi(r)/dr$, and
$H^{\pi}(r)=\sum_{k=1}^3h^\pi_k\,Y(\mu^\pi_k,r),$ 
where the constants are given by Eqs. (\ref{mas}), (\ref{const}). We denote the spin-coupling
$[\sigma_1\otimes\sigma_2]_{\rho,m}$ by $\Sigma_{12}^{\rho,m}$.

\medskip

2.{\underline {$\ \ \pi$-Exchange Current}}

The multipole for the true-$\pi$ exchange is the sum
\begin{equation}
T_{JM}^L\ =\ T_{JM}^{L,X1}\ +\ T_{JM}^{L,X2}\ +\ T_{JM}^{L,X3}\nonumber
\end{equation}
where
\begin{eqnarray}
\label{c2}
T_{JM}^{L,X1}\ =\ -{\frac {\sqrt{4\pi}} {i^{J-\epsilon}}} {\frac {4}{\pi}}
\left({\frac {f_0}{m_\pi}}\right)^2\,
\sum_{\ell \rho \sigma \sigma^\prime}
\sum_{L^\prime {\cal L}}\,i^{\sigma+\sigma^\prime + 1}
(-1)^{\sigma}(\hat\ell)^2 \hat\rho \hat L\ (\hat L^\prime)^2 \hat\sigma
\hat\sigma^\prime\hat{\cal L}\nonumber\\
\ \left(\begin{array}{ccc}1&1&\rho\\0&0&0\end{array}\right)
\left(\begin{array}{ccc}\sigma^\prime&\ell&L\\ 0&0&0\end{array}\right)
\left(\begin{array}{ccc}1&\ell&L^\prime\\ 0&0&0\end{array}\right)
\left(\begin{array}{ccc}L^\prime &\rho&\sigma\\ 0&0&0\end{array}\right)
\left\{\begin{array}{ccc}L^\prime&\rho&\sigma\\{\cal L}&\sigma^\prime &J\end{array}\right\}
\left\{\begin{array}{ccc}1&\ell&L^\prime\\ \sigma^\prime&J&L\end{array}\right\}
\nonumber\\
\  j_{\sigma^\prime}(qz)
\  \Phi_{\sigma,\ell}^{(3)}(q,r)
\ \left[[Y_{\sigma^\prime}({\hat{\bf z}})\otimes Y_\sigma({\hat{\bf r}})]^{\cal L}
\otimes \Sigma_{12}^{[\rho]}\right]_M^J\,\,(\tau_1\times\tau_2)_z\ ,
\end{eqnarray}
\medskip
\begin{eqnarray}
\label{c3}
T_{JM}^{L,X_2}\ =\ {\frac {\sqrt{4\pi}}{i^{J-\epsilon}}}{\frac{q^2}{\pi}}
\left({\frac{f_0}{m_\pi}}\right)^2 \,
\sum_{\ell \rho \sigma \sigma^\prime}
\sum_{L^\prime {\cal L}}\,i^{\sigma+\sigma^\prime + 1}
(-1)^{\sigma^\prime}(\hat\ell)^2 \hat\rho \hat L\ (\hat L^\prime)^2 \hat\sigma
\hat\sigma^\prime\hat{\cal L}\nonumber\\
\ \left(\begin{array}{ccc}1&1&\rho\\ 0&0&0\end{array}\right)
\left(\begin{array}{ccc}\sigma^\prime&\ell&L^\prime\\ 0&0&0\end{array}\right)
\left(\begin{array}{ccc}1&\ell&\sigma\\ 0&0&0\end{array}\right)
\left(\begin{array}{ccc}L^\prime &\rho&L\\ 0&0&0\end{array}\right )
\left\{\begin{array}{ccc}L&\rho&L^\prime\\ {\cal L}&1 &J\end{array}\right\}
\left\{\begin{array}{ccc}1&\ell&\sigma\\ \sigma^\prime&{\cal L}&L^\prime\end{array}
\right\}\nonumber\\
\  j_{\sigma^\prime}(qz)
\  \Phi_{\sigma,\ell}^{(1)}(q,r)
\ \left[[Y_{\sigma^\prime}({\hat{\bf z}})\otimes Y_\sigma({\hat{\bf r}})]^{\cal L}
\otimes \Sigma_{12}^{[\rho]}\right]_M^J\,\,(\tau_1\times\tau_2)_z\ ,
\end{eqnarray}
\medskip
\begin{eqnarray}
\label{c4}
T_{JM}^{L,X_3}\ =\ -{\frac{\sqrt{4\pi}} {i^{J-\epsilon}}}{\frac{4\sqrt{3}q}{\pi}}
\left({\frac{f_0}{m_\pi}}\right)^2\, \sum_{\ell f \sigma \sigma^\prime}
\sum_{L^\prime J^\prime {\cal L}}\,i^{\sigma+\sigma^\prime + 1}
(-1)^{{\cal L}+1}(\hat\ell)^2 \hat L\ (\hat L^\prime)^2 \hat\sigma
\hat\sigma^\prime\hat{\cal L}(\hat{J^\prime})^2 (\hat{f})^2\nonumber\\
\ \left(\begin{array}{ccc}1&\ell&J^\prime\\0&0&0\end{array}\right)
\left(\begin{array}{ccc}1&J^\prime&\sigma\\ 0&0&0\end{array}\right)
\left(\begin{array}{ccc}L&1&L^\prime\\ 0&0&0\end{array}\right)
\left(\begin{array}{ccc}L^\prime &\ell&\sigma^\prime\\ 0&0&0\end{array}\right)
\left\{\begin{array}{ccc}L^\prime&\ell&\sigma^\prime\\ \sigma&{\cal L} &f\end{array}\right\}
\left\{\begin{array}{ccc}1&\ell&J^\prime\\ \sigma&1&f\end{array}\right\}\nonumber\\
\ \left\{\begin{array}{ccc}L^\prime&f&{\cal L}\\ 1&1&1\\ L&1&J\end{array}\right\}
 j_{\sigma^\prime}(qz) \  \Phi_{\sigma,\ell}^{(2)}(q,r)
\ \left[[Y_{\sigma^\prime}({\hat{\bf z}})\otimes Y_\sigma({\hat{\bf r}})]^{\cal L}
\otimes \Sigma_{12}^{[1]}\right]_M^J\,\,(\tau_1\times\tau_2)_z\ .
\end{eqnarray}
Here $\Phi_{\sigma,\ell}^{(n)}(q,r)=\sum_{k=1}^3h^\pi_k\,\phi_{\sigma,\ell}^{(n)}(q,r,\mu^\pi_k)$,
where the functions $\phi_{\sigma,\ell}^{(n)}(q,r,m)$ are defined in \cite{fa}.
The multipoles (\ref{c1}--\ref{c4}) are real. 

\medskip

3.{\underline {$\ \ \rho$-Exchange Currents}}

The multipoles of the $\rho$-exchange currents can be obtained from
the above $\pi$-exchange currents by means of the following
replacements: $\mu^{\pi}_k\to\mu^{\rho}_k$, $h^\pi_k\to h^\rho_k$, 
\begin{equation}
\left(\frac{f_0}{m_\pi}\right)^2\ \to\ \frac{1}{4\pi}\left(\frac{g_\rho}{2M}\right)^2
\left(1+\frac{f_\rho}{g_\rho}\right)^2,
\end{equation}
and by inserting in each equation above the factor
\begin{equation}
-6 (-1)^\rho \left\{\begin{array}{ccc}1&1&1\\ 1&\rho&1\end{array}\right\}.
\end{equation}

\newpage

\centerline{CAPTIONS TO FIGURES}

\begin{figure}[ht]
\caption{Effects of the various contributions on $R_T$ 
in the quasi-elastic region at $q$=174 (upper left), 250 (upper right),
400 (lower left), and 500 MeV/c (lower right):
one-body (dotted), one-body + implicit MEC via Siegert operator (dashed), 
one-body + $\pi$-MEC + $\rho$-MEC
(dashed dotted), one-body + $\pi$-MEC + $\rho$-MEC + additional 
MEC via Siegert operator (solid).
}
\label{fig1}
\end{figure}

\begin{figure}[ht]
\caption{As Fig.1 but for the high-energy region.}
\label{fig2}
\end{figure}

\begin{figure}[ht]
\caption{Comparison of theoretical and experimental $R_T$
at $q$=250 (upper panel), 400 (middle panel), and 500 MeV/c (lower panel).
Theoretical $R_T$ with contributions: one-body (dotted) and one-body + $\pi$-MEC +
$\rho$-MEC + additional MEC via Siegert operator (solid). Experimental data from 
\cite{Saclay} (triangles), \cite{Bates} (circles), and \cite{world} (squares).}
\label{fig3}
\end{figure}
\nopagebreak
\begin{figure}
\caption{Comparison of theoretical and experimental low-energy $R_T$
at $q$=0.882 (upper panel), 1.64 (middle panel), and
2.47 fm$^{-1}$ (lower panel). Notation of curves as in Fig.3, but
additional curve in upper panel for total result in case that Coulomb force is neglected
in the final state interaction (dash-dotted).
Experimental data from \cite{Retzlaff}.}
\label{fig4}
\end{figure}

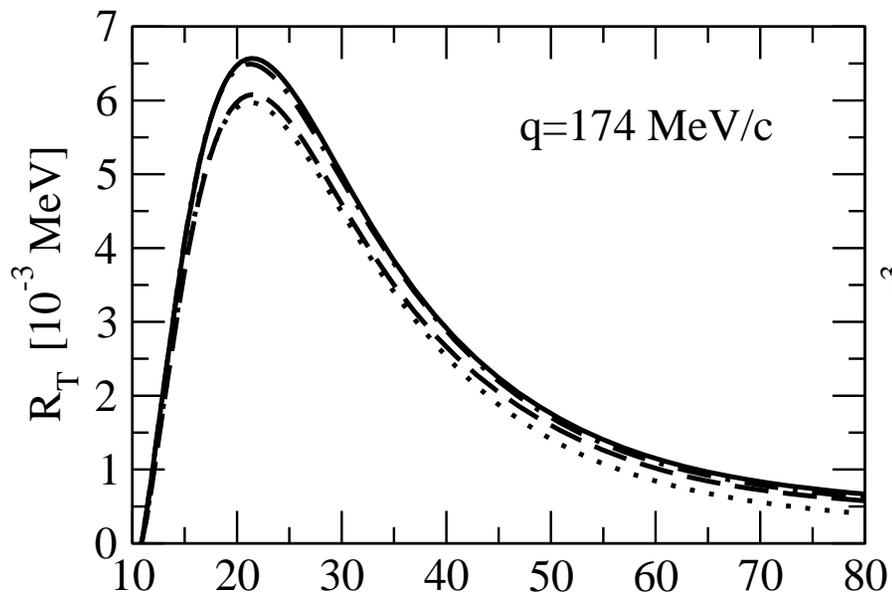
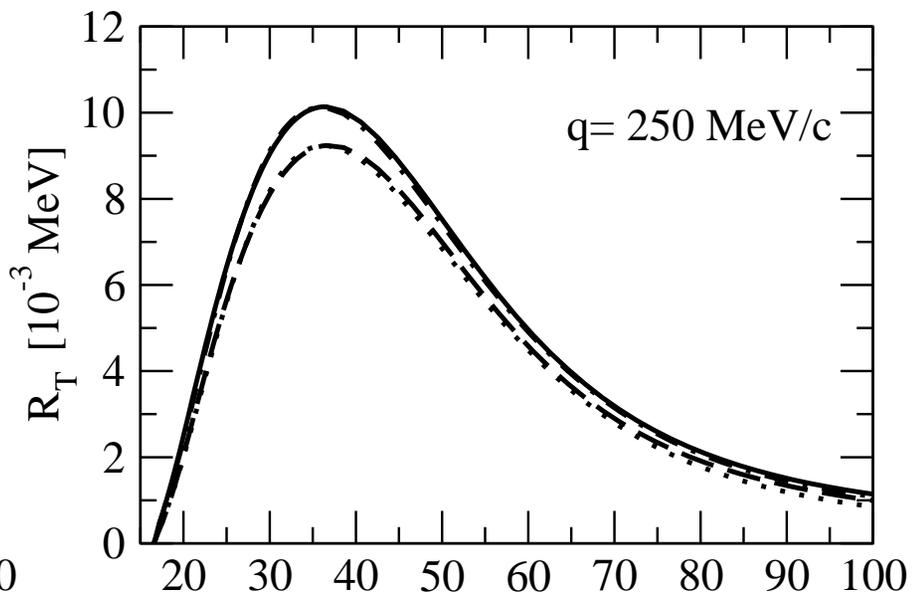
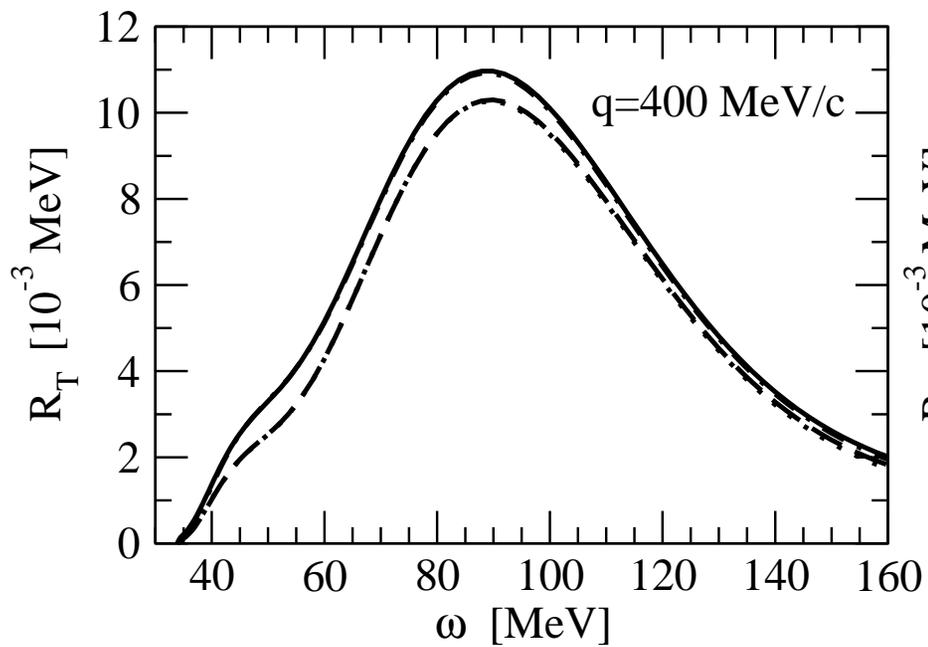
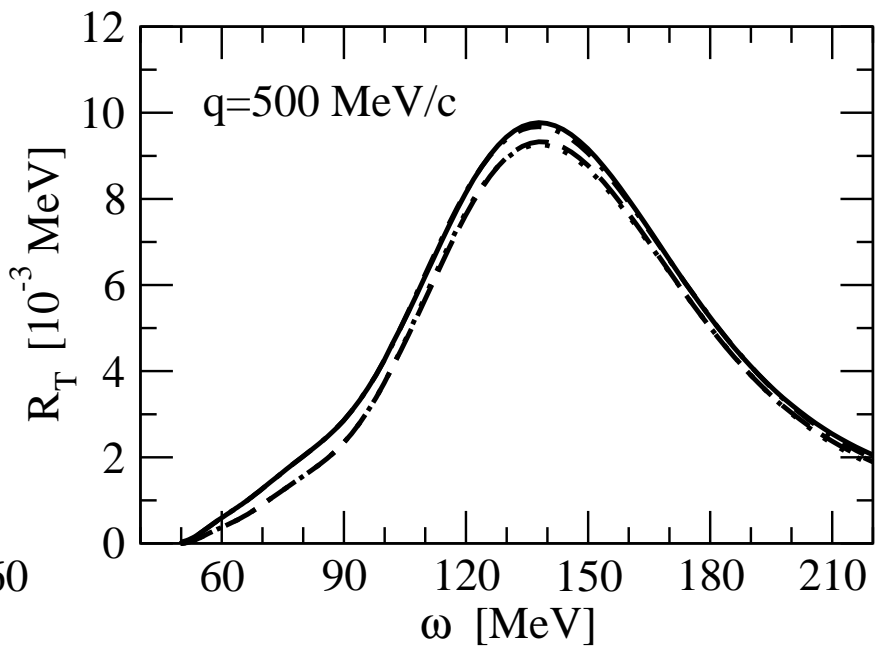
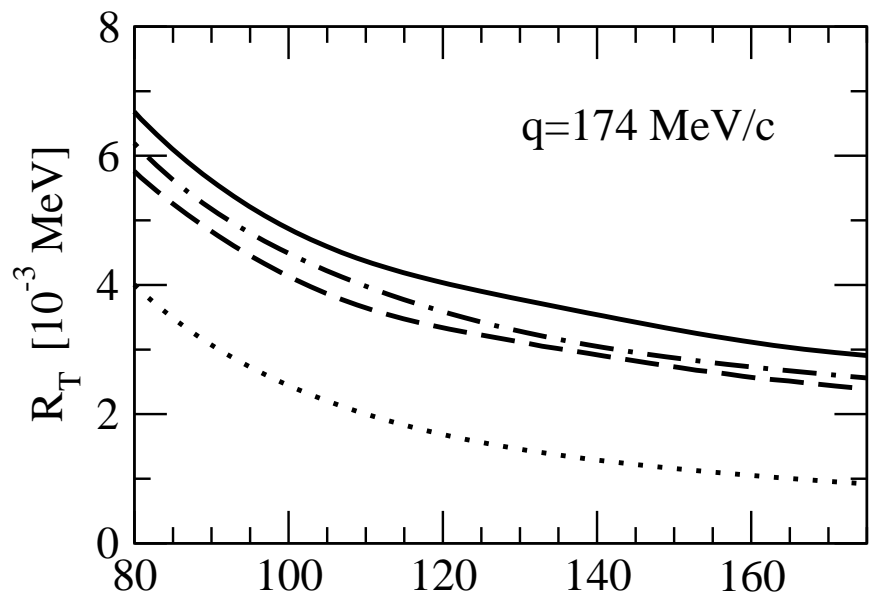
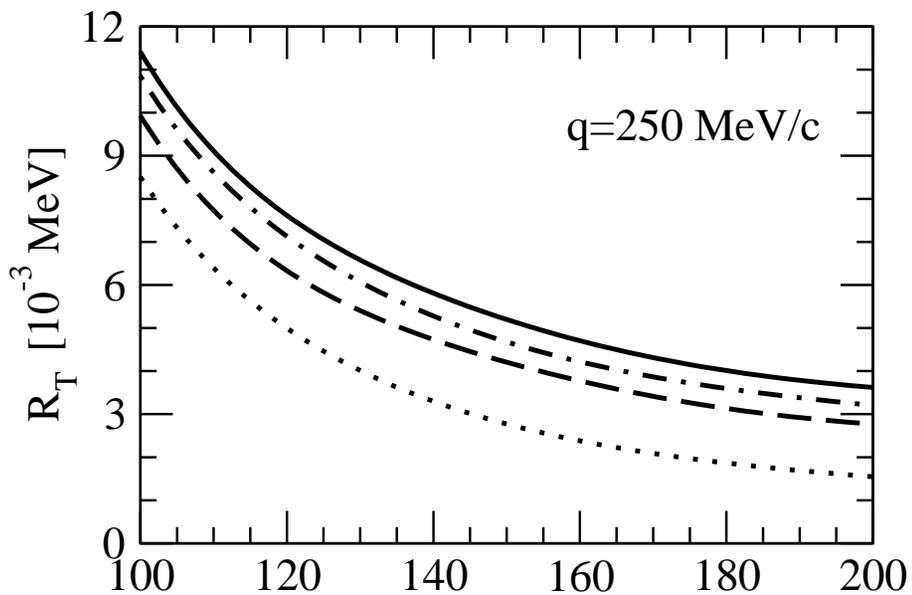
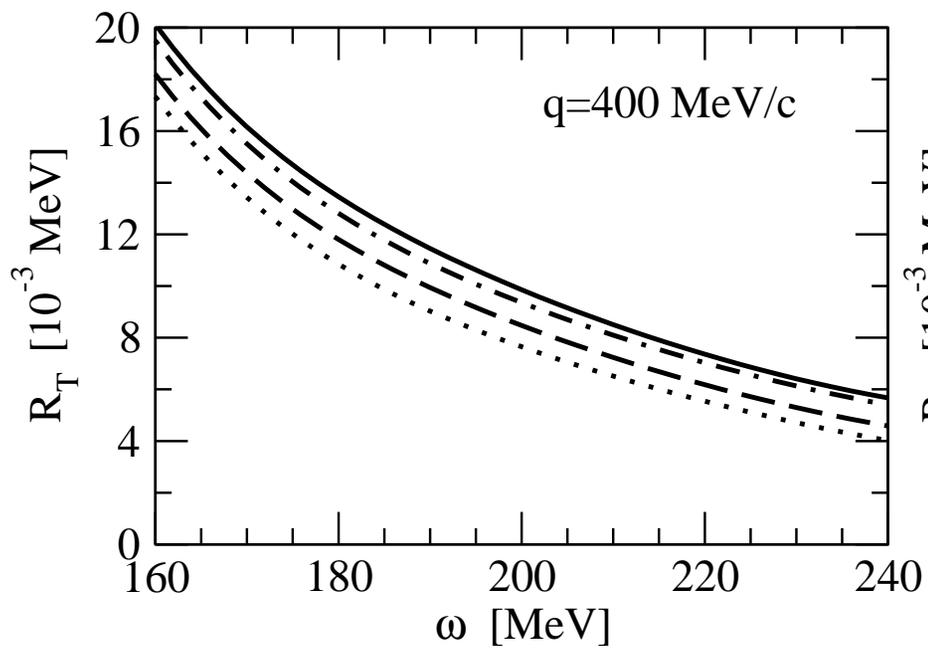
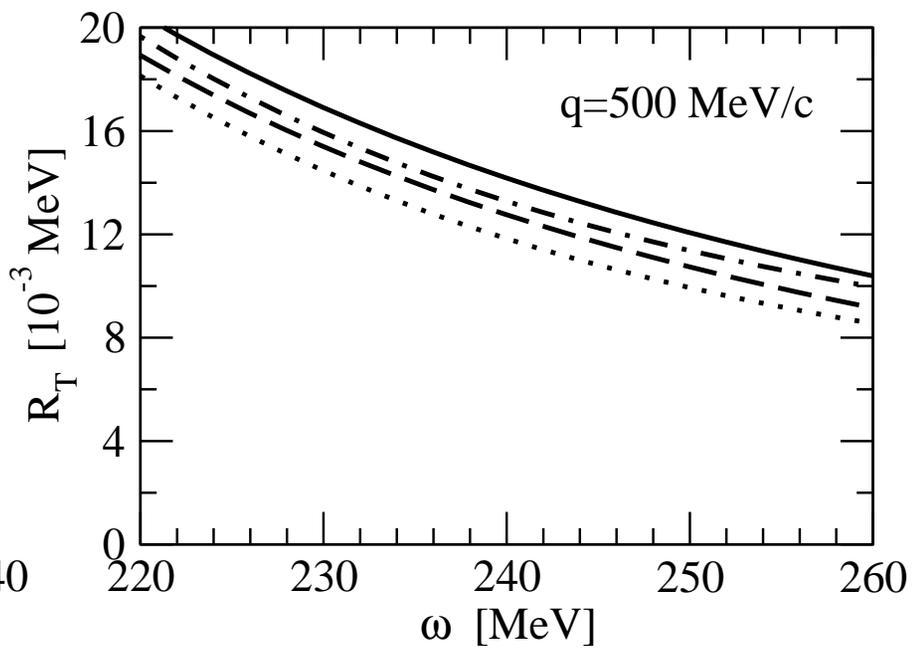
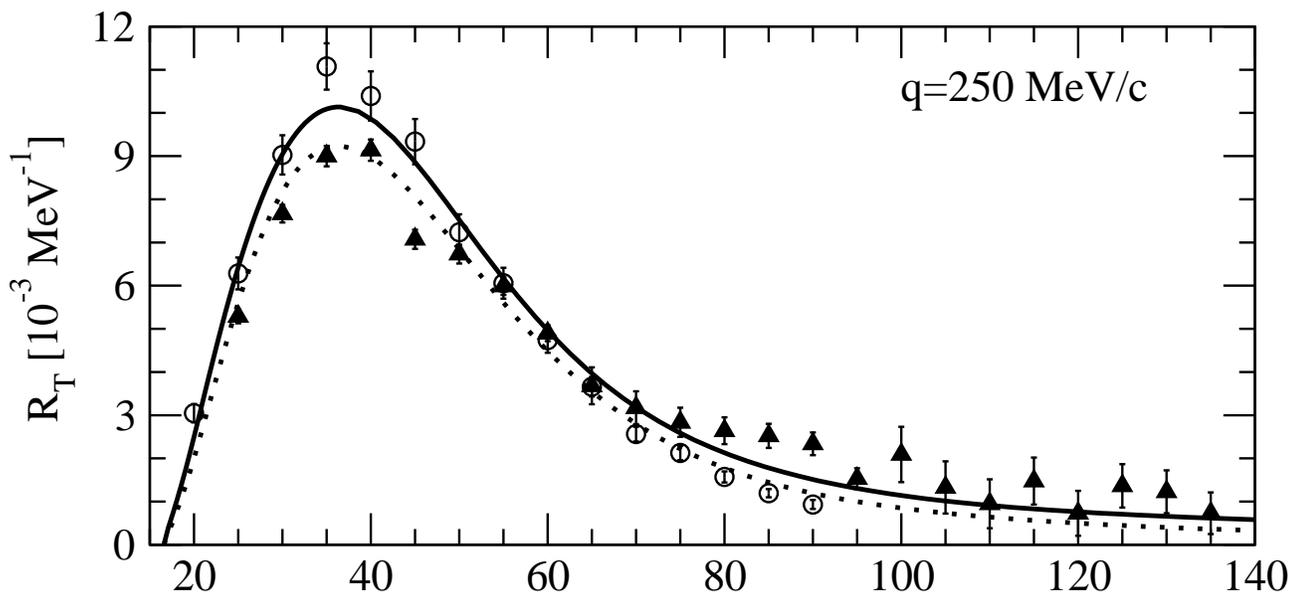
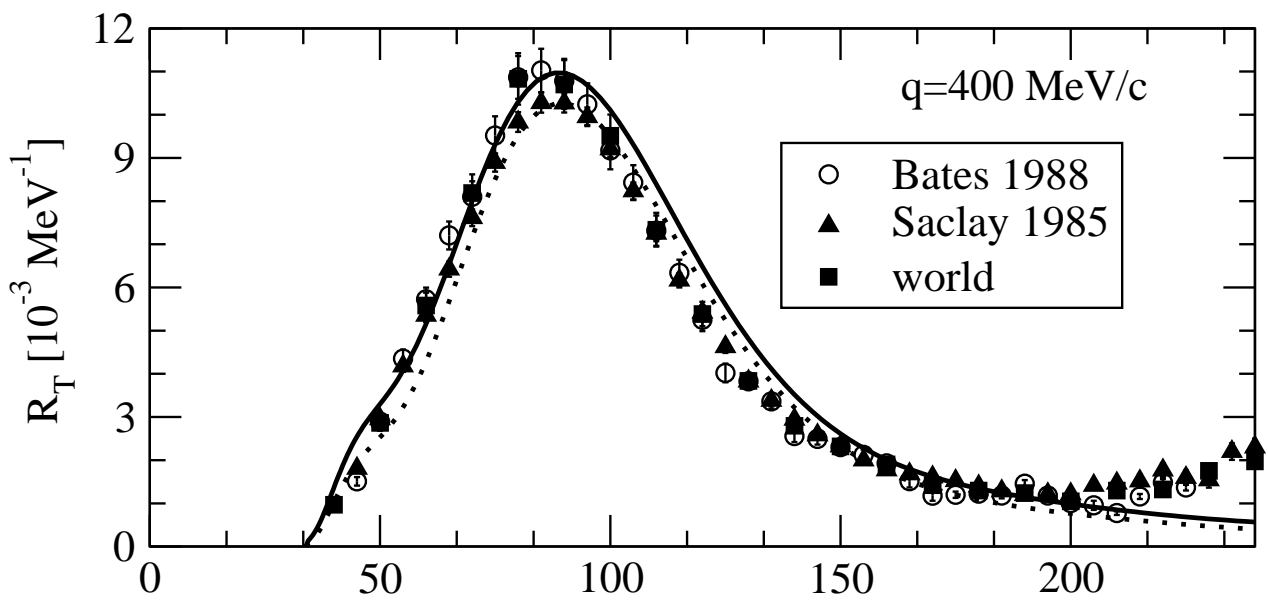
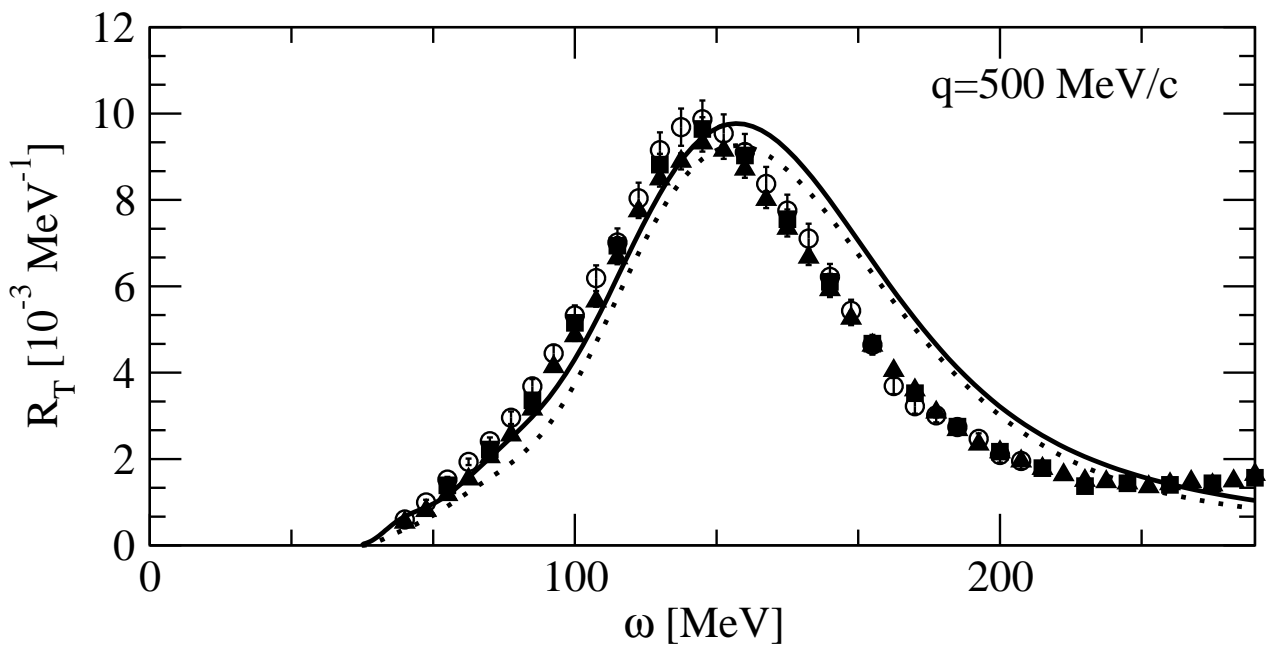
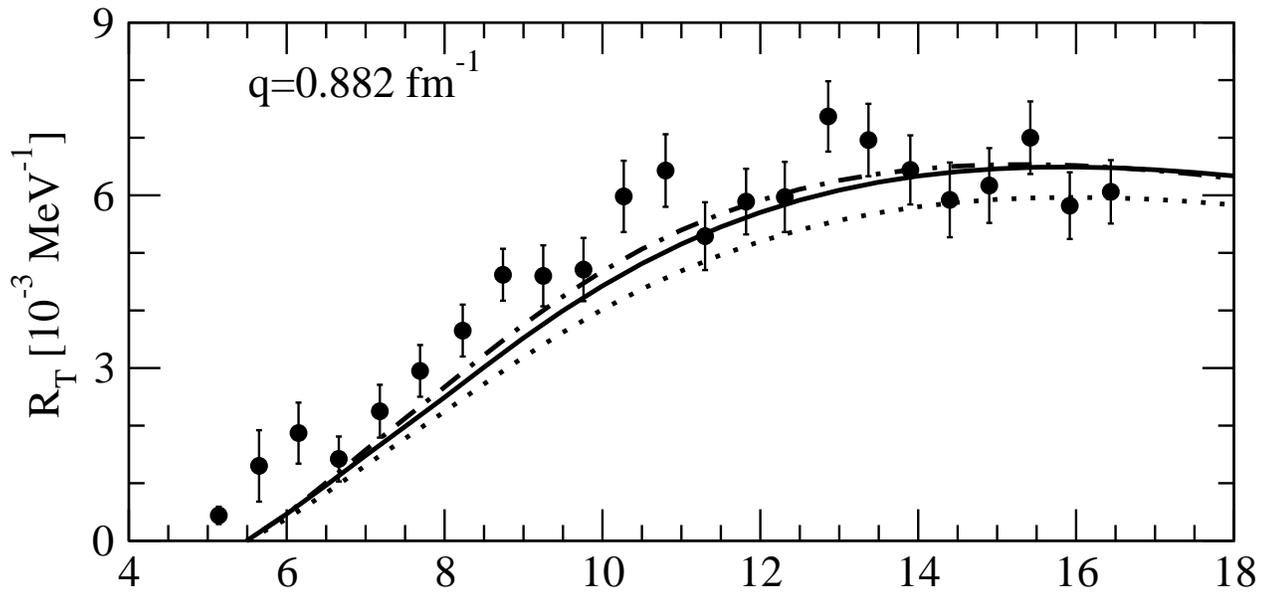
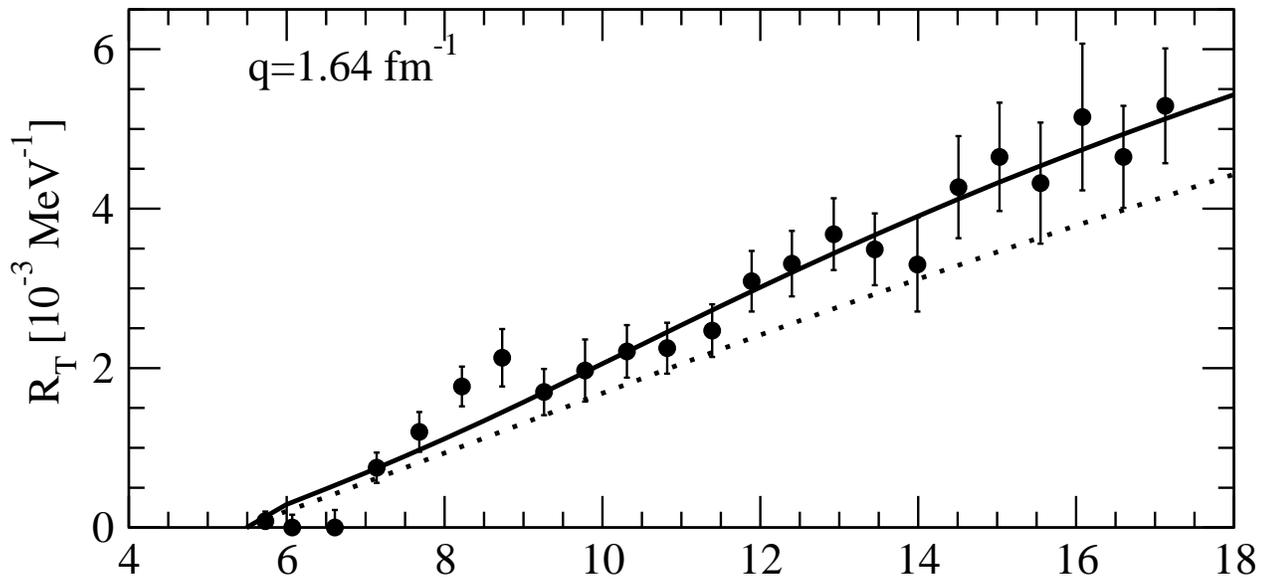
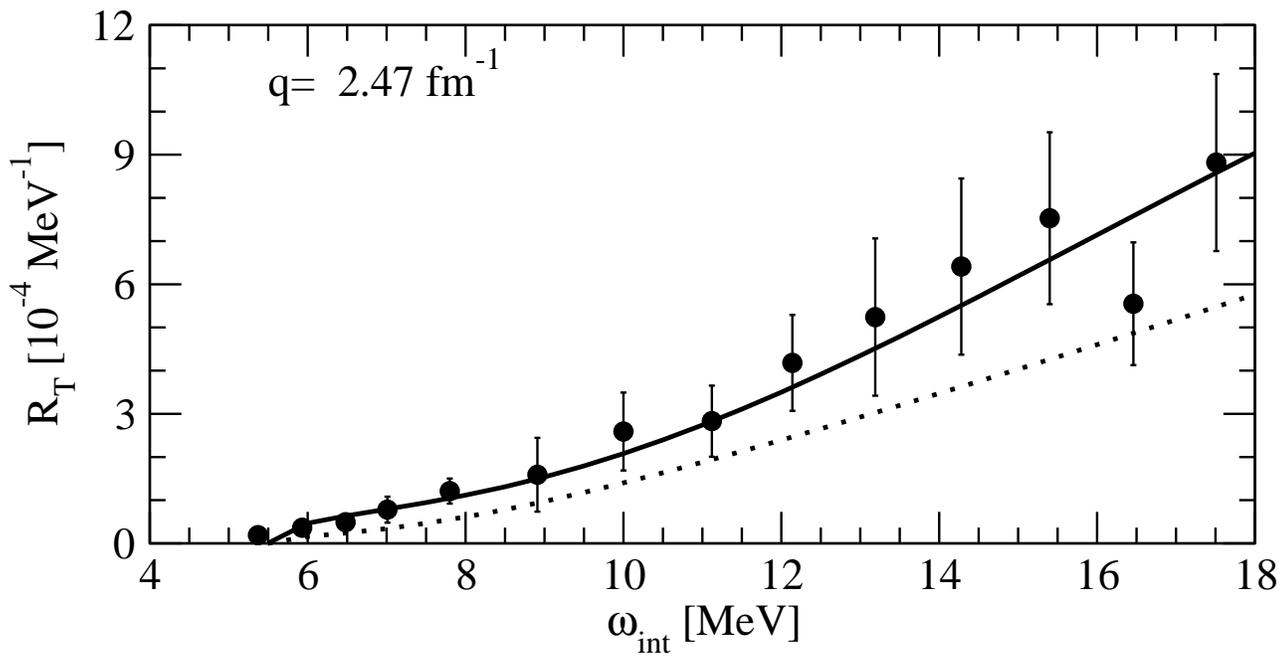
\end{document}